# Lessons Learned from Efforts to Standardize Streaming In SQL

*Authors: Sabina Petride, Dan Sotolongo, Jan Michels, Andrew Witkowski, Cara Haas, Jim Hughes*

## 1. Introduction

Under the logo "Where IT all begins", INCITS (InterNational Committee for Information Technology Standards [12]) oversees the development of standards broadly pertaining to information technology. For the database community, this is the organization known in the US to oversee the publication of the SQL standard [26], its revisions and most notably its extensions. Extending a standard with high outreach such as SQL is a lengthy, technically and logistically complex process, for obvious reasons. Any extension to the language must prove relevant to the industry. The extension must be vetted by representatives of the industry, which requires mutual understanding of the subtle differentiator points, ability to reach compromises, and a very good documentation process. Lastly, the marathon of revisions needs to be run to the finish line.

In recent years, the SQL standard has been extended, for example, to allow JSON data processing [18] and query property graphs, all increasingly popular. There is little debate that another very popular trend is streaming data from various sources for the purpose of analysis, summarization, or real-time notification ([6], [7], [8], [9], [22], [27]). SQL-like languages are used for processing data that is streamed ([1], [4], [5], [14]). Vendors introduce their own SQL dialects and explain the streaming specifics to audiences already well-versed in SQL, though semantics differences exist and are often a source of confusion (see [3], [10], [12], [13], to name just a few). Often, streaming data or data that results from its processing must be queried together with data traditionally stored in a database or materialized for further processing.

Acknowledging the reality of streaming languages and platforms overlapping with SQL and database systems, in 2019 INCTIS Data Management established an Expert Group with the focused mission to initiate the process of standardizing streaming support in SQL. Over time, the roster included companies like Actian, Alibaba, Amazon Web Services, Confluent, dbt Labs, Google, Hazelcast, IBM, Materialize, Microsoft, Oracle, Snowflake, SQLstream and Timeplus.

For the span of more than one year, representatives of each company have presented key features of their streaming product or, in some cases, multiple streaming products. These were live technical Q&A sessions accompanied by summary or position papers, which are unquestionably valuable. As expected, substantial time was spent in clarifying what common terms meant in each system and setting up a glossary. These sessions were followed by clarification notes and debates, and decisions that appeared mandatory to allow further progress. This first phase was followed by the next phase, which consisted of the group meetings, in which the expert group (EG) agreed on main exit criteria topics that a streaming solution in SQL must address, and position papers and follow-ups were written and discussed. This paper summarizes these group efforts, up to the summer of 2023.

While EG has identified a common core of streaming SQL topics that warrant an extension, the committee felt the modern streaming SQL space is too nascent to establish a complete set of abstractions that addresses each and all these topics coherently. No existing solution has yet proven the feasibility and utility of all such abstractions. EG shares the problems here, such that discussion may be resumed upon further validation of these ideas. It is fair to say upfront that no unified change proposal has been officially submitted for general revision as a result. There were numerous challenges to the production of such a proposal, including a pandemic, geographically large spread of attendees, member company switches, and the almost impossible task of balancing development-heavy tasks with writing and debating duties. However, EG believes all these hurdles are likely inevitable, and foreseeable, but most importantly surmountable if similar efforts are revamped. Ultimately, EG concluded that the standardization of streaming SQL requires more time for the community to learn from the variety of solutions on the market. In the meantime, this group shares its learnings, since the value of standardizing streaming in SQL remains unchanged, if not higher than when EG meetings started.

# 2. Simple Examples – All Main Subtleties Arise

It is interesting to witness what happens when SQL-like statements often executed in streaming systems are discussed in a group that mixes streaming practitioners and savvy tech developers together with SQL veterans. Take for instance the simple example

SELECT * FROM myStream;                                                                                                    (a)

Let's for a moment accept that myStream is a source of relational data but agree not to label it yet as either a table or a new type of object.

A SQL user expects the statement to either return no rows or return some set of rows, and eventually for the control to come back to the user. It may take arbitrarily long to return the first row or all the rows, but the control will be returned to the user. The rows represent all the data in the myStream source as of the current time, given the transactional context we are executing the statement in. The result is deterministic under a fixed context – if the statement is executed a $2^{nd}$ time, it will return the same set of rows.

In a streaming system, the statement is not expected to end. The user may see no rows or several rows, and later more rows are returned, and in theory there may always be more rows to be returned. The statement must be interrupted/aborted for it to end. If the statement is executed a $2^{nd}$ time, it may not return the same rows. One intuition is that the user subscribes to myStream, or rather changes to it, and passively receives rows, and if interest is lost in the stream, then the user cancels the subscription. We can already see why these semantics can be unsettling for SQL veterans but makes perfect sense for streaming applications.

Some features of myStream are assumed implicit:

- It has a notion of processed/old vs. unprocessed/new rows.

- It recalls what has been processed so far.

- Rows are always added to it.

Some systems explain a stream as a book with a bookmark and expect the user to always read from the bookmark and advance it during reading. Some systems offer knobs to be turned to control the batch size of events/rows that must arrive new in the source before they are returned to the user – we can think of this as reading reasonably sized paragraphs at a time vs reading each word and then waiting for the next one to appear.

Are these properties inherently tied to myStream however? Can these properties be rather associated with a cursor for SELECT * FROM myStream; instead? Afterall, likely there are concurrent users that execute the same or similar statements all involving myStream. If sharing a book while reading, readers can be made responsible for their own bookmark.

Can we have a language that allows us to choose when we want to look at myStream as streaming systems do and when we want to look at it as current SQL would do?

We can frame this question as a stream vs table dilemma: Is myStream a new type of object, call it stream, whose presence in a statement affects the semantics of the statement? Or is myStream just a table, perhaps with additional constraints such as rows are never updated, only inserted/appended. In the latter case we need additional syntax to distinguish when we would like to look at myStream and see only newly appended rows, and when we want to look at myStream as a table and see all its data. After all, even in simple analogies like a book with a bookmark, the same reader may be interested in reading back from arbitrary points, and then resume from the bookmark. EG discusses our solution for this dilemma in the Stream Or Table section, while in section Continuous Cursors and Subscriptions EG summarizes the main proposal for differentiating between the two expected behaviors with new syntax.

Assuming myStream has a price column, let's consider a statement like

SELECT AVG(price) FROM myStream;                                                                                                (b)

Even if rows in myStream are not expected to be updated, the average price can change. Should the current average price be returned to the user as new rows are added to myStream? Should the current average price be returned only if it is *different* than the last average reported? We will call the latter as *delta-semantics* – only changes with respect to the previous result are returned.

Delta-semantics requires keeping some state – such as the last reported average. It may make sense to report changes more explicitly, such as when the average price changes from a1 to a2 to report (-, a1), (+, a2). That is 2 records: one indicating that the previous value was a1 and should be deleted, under the assumption that the returned values are materialized, and one record making explicit that the new value is a2 and it should be added to the state.

Perhaps the user can have control over the format. One way to achieve that is to introduce a SQL relational operator, CHANGES, that reports the delta changes to its input relation relative to last invocation, along the lines of

SELECT * FROM CHANGES(SELECT AVG(price) FROM myStream);

Aggregated streaming data is often further processed, either in a streaming way or not. The language needs to allow materialization of streaming results into relations. We refer to this as *streaming, or continuous, DML*. The format of the data that is materialized in a streaming way is tightly related to the output of the CHANGES operator, if we choose to make it explicit, or to the internal format in which changes to a previous result are expected to be represented to be persisted.

For our example, assume that we would like to materialize the average price from myStream into a table myDestination. If we believe that the format for representing changes to the average is not relevant to the user, then we can think of the operation as keeping myDestination in sync with

SELECT AVG(price) FROM myStream;

so that at each point in time

SELECT * FROM myDestination;

would return the most up to date average price. This is akin to myDestination containing the same data as a materialized view for SELECT AVG(price) FROM myStream;

and continuously refreshing the materialized view.

If we believe that there is value in exposing the change format, then we can expect users to run a DML statement that applies rows from

CHANGES(SELECT AVG(price) FROM myStream)

to myDestination. Applying changes for materialized view refresh is however not exactly a MERGE statement that would delete '–' delta records and insert or update '+' delta records. It requires ordering change records in time order or commit sequence number, for instance, and applying changes in that order.

Regardless of whether a new form of DML statement is best suited, or whether CHANGES operator is to be used explicitly, several other questions easily arise. In streaming systems, we often refer to the arrival of events and processing that is triggered by the arrival of events, or of a batch of events. This is not a notion known in SQL. Instead, we would need to rephrase the semantics in terms of INSERT time or COMMIT time. For example, instead of expecting myDestination to be refreshed on arrival of events/rows in myStream, we could expect myDestination to be refreshed on COMMIT of a transaction that affects myStream.

Next, there seems to be a need to formalize in SQL the notion of scheduling of operations, such as refresh of a target table based on a statement, and the schedule likely involves CRON-like expressions and reference to COMMIT of relevant transactions. These topics are detailed in the section Streaming DML, though the group also debates whether scheduling should be a vendor option only, not mandated by the SQL standard.

One can easily argue that typical streaming systems are not likely to compute a running aggregation over an entire stream, but rather group together events/rows in the stream by some time dimension or other dimension that is mostly increasing with time, and report aggregations relative to a group. These groups are mainly called *windows*, and come in different flavors, depending on how we choose to formalize their starting point and range ([5], [14],[16], [17]). Typical windows are defined on the time dimension of a stream data source, and they are either of fixed range not overlapping (typically called *tumble windows*), or fixed range overlapping (*hopping windows*), or session-windows, where the grouping is more dynamic. For instance

SELECT AVG(price) FROM myStream

WINDOW HOPPING (SIZE 30 SECONDS, ADVANCE BY 10 SECONDS); (c)

is expected to logically partition the rows in myStream on their time attribute, one group for the range 0 to 30 seconds, a 2$^{nd}$ group for the range 10 seconds to 40 seconds and so on and compute the average price for each group. While popular, these statements can be seen as incomplete or ambiguous to a SQL user, for a couple of reasons:

- WINDOW here is not the SQL WINDOW

- There is an implicit grouping on the window identifier, while the statement does not consistently contain GROUP BY, across streaming products offerings.

- There are hopping, tumbling and session windows, which seem like syntactic sugar for special predicates that define the bounds of each window. The obvious question is why these windows and not others, though this is a secondary concern.

- There is an implicit notion of time attribute of each stream and that it is mostly monotonically increasing. In SQL we would need to make this attribute identified by a user specified name. More importantly, in SQL there is no constraint for monotonicity or approximate monotonicity yet.

- There is an implicit notion of start or minimum time.

Though less of a point of focus in the group meetings, considerations on supporting time-windowing in SQL are detailed in sections Increasing Constraint – Definition, Usage and Generalization and Finalization.

If instead of returning all rows the statement returns aggregated data over time, then results will tend to change less often. However, aggregated values can still change. For example, in theory, if rows in myStream can arrive arbitrarily late, then it is always possible that a correction to the last reported average price for say time-window midnight to 30 seconds past midnight today will be required, so the average for this window should be persisted in the state for the statement. For that matter, the average for each time window should be persisted, which can result in arbitrarily large space requirements. In practice however, this is unlikely to be necessary. There is a need to properly formalize in SQL what constitutes a *late* row – with respect to a time dimension.

Once a proper definition of late rows exists, the question of handling late rows remains. One can choose to ignore late rows or to materialize them in some side table. One can choose to report data even if late rows may still be possible to occur, or, on the contrary, to withhold from reporting rows of a time-window until all rows that are still possible to be added to myStream would be guaranteed to be declared late. In the former case, special syntax has been invented such as EMIT CHANGES, EMIT FINAL to indicate the desired behavior. In the latter case, systems have relied on environment settings – such as at session or database level - to inform when events are guaranteed to be declared late and ignored, so that aggregates over some time-windows can be finally reported. One observation arising from the analysis of such ad-hoc solutions is that there is a need to formalize a notion of *finalization* – a condition that determines when some result set is guaranteed not to change when evaluated later. A proposal has been made to formalize *finalization on write* – as a new form of constraint to be enforced at DML or COMMIT time. For example,

ALTER TABLE myStream

*FINALIZE (time <= time_constant)*;

 would ensure that no record with time attribute at most time_constant is allowed to be inserted into myStream. Time-windows with upper bound below time_constant are therefore known as not susceptible to late events and therefore they can be reported in a statement that explicitly requires only final results, like

*FINAL*(
    SELECT AVG(price) FROM myStream
    WINDOW HOPPING (SIZE 30 SECONDS, ADVANCE BY 10 SECONDS));

 When a write-time constraint is undesirable, finalization can be specified as part of queries using *finalization on read*. This would apply to cases where there can be no control over the source data, but at statement level we could ignore rows that do not satisfy the finalization condition.

FINAL(
    SELECT AVG(price) FROM myStream
    WINDOW HOPPING (SIZE 30 SECONDS, ADVANCE BY 10 SECONDS)
    FINALIZE (time <= time_constant));

This is briefly discussed in the Finalization section.

SQL as a query language is not concerned with the size of relations or in general with the performance of any systems that implement it. In principle, immutable data may not be deleted—it grows forever, often at a high rate. However, many streaming applications have no use for very old data. For example, time-window statements with some form of finalization would not scan data for already reported time-windows. To a minimum, there is therefore scope for information life cycle policies for streaming data sources that are aware of the workload – such as automatically moving cold data to a space efficient formatting, as determined based on time attributes and workload. There is therefore a concern that SQL should allow users to declare data *expiration* conditions. Much like finalization, expiration policies might make sense as uniform lifecycle policy based on a condition, such as

ALTER TABLE myStream

EXPIRE (time <= time_constant);

Expiration gives implementations the freedom to free the storage associated with these rows, but the exact semantics can be debated. For example, expired rows may or may not be considered in computations. Implementations may actually purge or move to different storage expired rows. These processes can happen at different times. More complex aspects of expiration are discussed in section Expiration.

# 3. Stream Or Table?

In the initial debates there were strong believers that streaming data sources are not tables or relations as defined in SQL. However, discussions matured around operations allowed on streaming data sources, such as subscription-based read, expiration and finalization. It eventually became clearer that a lot can be achieved if streaming data sources are seen just as regular SQL relations with additional properties enforced as either new type of constraints or via new ALTER statements.

Typically, streams are described as continuous flows of events – new events arrive continuously or under some regular pattern. As we mentioned above, arrival of an event in streaming systems is akin to INSERT of a row in SQL. Therefore, streams can be modeled as SQL relations in which rows can only be inserted. EG calls such relations *insert-only* and discusses them in Section Insert-Only Relations.

Streaming sources are however just a particular case of insert-only relations. Any change log on top of an arbitrary SQL relation is also insert-only. Exposing the change log as a relational operator is discussed in Section CHANGES SQL Operator.

## 3.1 Insert-Only Relations

EG defines a relation T to be insert-only if and only if, for all times $t_1 < t_2$, the set of rows in T at time $t_1$ is a subset (not necessarily strict) of the set of rows in T at time $t_2$.

If T is a persistent table, a new type of constraint, called INSERT ONLY, was suggested. It guarantees that T is insert-only as defined above. The syntax could be as simple as

```
CREATE TABLE T <table_definition_clause> INSERT ONLY;
```

for creating an insert-only table, and

```
ALTER TABLE T INSERT ONLY;
```

for altering an existing table as insert-only.

In practice, one way to guarantee that a persistent table is insert-only is to raise errors when any DML other than inserting rows is attempted on the table.

However, EG would like to reason about non-persistent tables (also known as the result of query expression) as insert-only as well. For instance, an equi-join on insert-only relations is easily seen to be insert-only. For this purpose, EG finds it convenient to define a set of syntactic rules for proving that a relation is insert-only. The complexity of SQL syntax makes it difficult to prove the completeness of such rules, and EG leaves it to future work to expand the rules beyond the list below, based on [26] section 4.14:

- <with_clause> is insert-only if all <query_expression>s and its <with_list> are insert-only

- <query_term> is insert-only if all its <query_primary> is insert-only

- INTERSECT is insert-only if all its <query_term>s are insert-only

- <query_expression_body> is insert-only if all its <query_term> are insert-only

- UNION is insert only if its -<query_expression_body> and <query_term>s are insert-only

- <query_primary> is insert-only if all its <simple_table>s and <query_expression_body> are insert-only

- a <simple_table> is insert-only if its <query_specification> is insert-only, and <table_value_constructor> is constant, and <explicit_table> is insert-only

- a <query_specification> is insert-only if <select_row> is a per-row constant or it is an insert-only <table_expression>

- a <table_expression> is insert-only if its <from_clause> is insert-only, its <where_clause>'s <search_condition> is monotonic – once TRUE it remains TRUE, it has no <group by> , <having> or <window> clauses

- a <table_reference> is insert-only if <table_factor> is insert only, and <joined_table> has both <table_reference> and <table_factor> insert-only and it is either a cross join, or an inner qualified join, or an inner natural join.

## 3.2 CHANGES SQL Operator

The SQL stream expert group has argued that incremental fast processing of changes in time to a dataset is at the heart of stream processing and that SQL is fundamentally missing a method to query the changes to a table or view, however. They subsequently contend that streaming SQL needs to be amended for this purpose.

The paper proposes augmenting SQL with a relational operator, called CHANGES, that returns the changes to its input relation. CHANGES can accept optional time range arguments, for filtering out the changes to the input relation that have occurred in the given time range. It can also accept optional format arguments, such as 'DELTA' or 'LOG'. Intuitively, 'LOG' corresponds to a change log, where each INSERT, UPDATE, DELETE operation on each row in the input relation is returned. 'DELTA' can consolidate the change representation so that redundant operations are removed (such as an INSERT followed by a DELETE of the same row).

- The result of the CHANGES operator may have a few additional columns relative to the input relation

A key feature of the result of CHANGES applied to any relation is that it is insert-only – the change log is only growing with time. Another important feature is that it is not limited to permanent tables – the changes over any query expression can be defined unambiguously. While we may be well accustomed to table logs and the distinctions between flavors of logs (such as supplemental or not), and their application in replication or recovery, CHANGES exposes the same level of logging information for any relation.

# 4. Continuous Cursors and Subscriptions

Recall that a SQL cursor is a mechanism by which rows of a relation may be acted on, one at a time. The most notable cursor behaviors are defined via 4 top-level properties:

- holdability: A WITH HOLD cursor can continue to be used outside of the transaction that has created it, while a WITHOUT HOLD cursor is transient to the transaction that has created it.

- sensitivity – For a WITH HOLD cursor, the sensitivity property controls whether changes made in transactions that have not created the cursor but that occurred while the cursor is open are visible during fetch from the cursor.

- scrollability – A SCROLLABLE cursor supports non-sequential fetches, such as FIRST, LAST, PRIOR.

- returnability – Returnability refers to cursors in the context of a stored procedure. WITH RETURN cursors are cursors whose result set table may be used as a result set from the procedure.

The EG discussed defining a *continuous cursor* as a cursor that returns rows from an *insert-only relation* and has the following properties:

- WITH HOLD and SENSITIVE, and only committed changes to its relations are made visible

- NO SCROLL – only NEXT/forward fetches are possible

- configurable returnability – either WITHOUT RETURN or WITH RETURN can be specified.

These properties together ensure that the initial FETCH from a continuous cursor returns all rows in the underlying relation, and each subsequent FETCH returns the newly committed rows in the relation.

Fetching a row from a cursor may return NO DATA status, such as when no new rows have been inserted into the underlying relation relative to the previous FETCH. It is an open question whether for continuous cursors it is beneficial to introduce a new NO DATA YET status.

Syntactically, the expert group would like to use a new CONTINUOUS token in the definition of a cursor, to specify that it should be treated as defined above. One option EG considered is to specify CONTINUOUS as a qualifier, preceding the token CURSOR, as in

```
CONTINUOUS CURSOR c is SELECT * FROM myStream;
```

It should be an error to define a continuous cursor on a relation that is not insert only, or to ALTER a permanent table to remove its INSERT ONLY property when there exist open continuous cursors depending, directly or indirectly, on the table.

Since insert-only deduction rules are not complete, it may be beneficial to syntactically restrict continuous cursors to

```
CONTINUOUS CURSOR C IS SELECT <select_list> FROM CHANGES(relation);
```

CHANGES operators on a permanent table known as INSERT ONLY can be omitted or rendered redundant.

Recall the simple example

SELECT * FROM myStream;                                                                                                         (a)

The above syntax will maintain its current SQL semantics and return all the rows in table myStream in the current transactional context. A new statement is required for streaming semantics, along the lines of

```
SUBSCRIBE TO

SELECT * FROM CHANGES(myStream);                                            (a - streaming semantics)
```

Internally, the above will define a continuous cursor

```
CONTINUOUS CURSOR c IS SELECT * FROM CHANGES(myStream);
```

(with a system generated name, here defined for simplicity as c), will OPEN the cursor and continuously FETCH from it. The control is not returned to the user. The user canceling the SUBSCRIBE statement or the session ending both result in closing the cursor. As mentioned above, if myStream is a permanent table with an INSERT ONLY constraint, CHANGES operator is optional, and the following syntax should be allowed:

```
SUBSCRIBE TO

SELECT * FROM myStream;                                            (a - streaming semantics insert-only table)
```

The above does not give the client any control on the timing of FETCH from the continuous cursor. Though CLI (client level interface) standardization is beyond the scope of this paper, EG acknowledges that such omission may not be seen as practical. One proposal is to allow syntax for specifying the FETCH schedule. This is akin to the schedule clause we detail in section 5.2.3. Omitting the schedule syntax simply implies that a default schedule is used (again, we can rely on the same default as for streaming DML). Another option is to leave the scheduling as implementation-defined.

# 5. Streaming/Continuous DML

## 5.1 What Is Continuous DML? What Problems does It Attempt to Solve?

A Continuous DML evaluates its underlying arbitrary query continuously and incrementally, if it is more performant, and applies the changes in the query result set to the target table.

Some examples are:

- SCDs (Slowly Changing Dimension tables) - In data warehousing, fact tables can be joined with dimension tables that change slowly, though possibly following an unpredictable schedule. Such joins are often materialized, for better query performance; if the materialization is up to date, then SQL queries can be internally rewritten against the materialization.

- Evaluate top 10 stock traders in a volatile dataset and materialize the results, for further/pipelined processing.

- Evaluate the total outstanding orders in an e-commerce platform, and, more generally, summarize volatile data through aggregation, and materialize the result into a table. Continuous cursors may be further defined on the target table, for different computations on these aggregated data.

- Evaluate security rules implemented in SQL, with timely actions triggered in case of rule violations.

Let's consider the case of SCDs. Several queries are based on the join between a fact table F and a dimension table D. F is mostly static and D is slowly changing. For simplicity, let's assume that F is not changing. The join of F and D is expensive and therefore materialized in table T, and queries are rewritten to scan T instead of recomputing F join D. Maintaining T in sync with F join D is in our opinion a form of continuous DML - with T as target and F join D as source relation/statement. The ideal behavior is for changes on D to be "immediately" reflected on T, which in the database world means that a COMMIT of any transaction affecting D must include a logical re-computation of T. For practical reasons, COMMIT should not be slowed down significantly, and therefore it is better for the re-computation of T to be incremental. Depending on implementation and on the volume of changes on D, there is a tradeoff between the ideal behavior and system performance. One mitigation method is to give more control to the user to formulate acceptable alternatives, such as

- instead of guaranteed correctness of T, allow changes on D to be reflected in T every S seconds or at a clear timely schedule; the user can change the schedule based on preference

- do not block the COMMIT of transactions on D before reflecting them on T, if the re-computation of T is guaranteed to occur on a best-effort basis after each COMMIT that affects D.

The EG's view is that we must have a clear default semantics together with syntax for most used schedules for keeping a relation/statement in sync with another relation.

EG should acknowledge that this is not a new problem at all - it is shared with systems that use streams to produce new streams, or rely on materialized views, as well as with those that implement stateful tasks to be executed in the database. Such systems all expect a schedule to be formulated for either task execution or materialized view table container refresh. Some of these systems implicitly expect some tasks to occur "on the arrival of an event" in any input stream. These are all non-standard SQL concepts which need translation in SQL. For instance, arrival of an event can be reformulated as COMMIT of an insert into an insert-only table, which could correspond to on INSERT for auto commit DMLs.

Back to the SCD case, once the schedule at which T is recomputed is clarified, we can focus on how the re-computation happens. Logically, with every change of D, T is recomputed, but a full re-computation is not always required, and it is often costly.

If it is known that D is an insert-only table, then based on our simplified assumptions F join D is also an insert-only relation. New rows of D, relative to the previous scheduled execution, that join with F can be inserted into T, as an incremental re-computation of T.

In practice, D is not guaranteed to be insert-only. Therefore, when rows are inserted into D, the ones that join with F should be inserted into T, and when rows are deleted from D then the ones that joined with F should be deleted from T as well. If a row in D is updated, UPDATE can be modeled by a DELETE followed by an INSERT, or it can be modeled as a special UPDATE change row. Incrementally re-computing T assuming access to change log records of D appears as a kind of MERGE statement, but writing a correct MERGE statement is not straightforward and it depends on

- LOG vs. DELTA change log formats

- how UPDATE is modeled

- assuming a timestamp column is set on the table, whether we expect timestamp value to advance with every COMMIT or whether we must also include a COMMIT sequence column in change log records

- whether row ids are maintained on UPDATE modeled as DELETE followed by INSERT, or whether a new row id is generated on the INSERT part of an UPDATE, and other minor details.

Note that continuous DML is not only referring to continuously materializing rows fetched from some continuous cursor. For instance, when composing continuous DML with triggers on the target table, it could be a powerful solution for use cases like evaluating security rules implemented in SQL. The trigger body may implement the actual action. In this case, continuous DML may be used just as a vehicle for triggering actions when changes to a statement/relation occur.

EG advises standardizing the following critical pieces of a continuous DML:

1. An easy-to-use high level SQL syntax.

2. When to trigger the underlying DML.

3. Where to pick up where it left since the last DML action: a continuous cursor. This avoids duplicate data / missing data, and it is tied to the behavior in case of errors.

We proceed by formulating some basic examples and use them to show some concrete challenges.

## 5.2 Proposed Syntax and Semantics

### 5.2.1 Basic Examples

#### 5.2.1.1 Change Log
Let's start with a very basic example. Recall that CHANGES is just a relational operator, with no materialization assumed. (Note that CHANGES has a RowID column, which is meant to contain a unique identifier for each row in the change log. However, it is not mandated that this row id should be obtained by materializing the change log. Our examples refer to the RowID column for ease of understanding only.)

For our example, we want to maintain table LogT as a materialized change log on T in DELTA format. To do so, we can run

```
2:00 > CREATE TASK q [ON COMMIT] AS
INSERT INTO LogT
SELECT CONTINUOUS *
FROM CHANGES(T, 'DELTA');
```

This is a special case of a new DDL for creating a new named SQL object, here called *task*, that fetches from a continuous cursor based on a schedule and inserts fetched rows into a target table:

```
CREATE TASK q [<schedule_clause>] AS
INSERT INTO <table|insertable view>
<continuous_relation> [<error_logging_clause>]
```

where

```
<continuous_relation> ::=
SELECT CONTINUOUS <select_list> FROM CHANGES(<relation>, <format>)
```

Here, relation is a table or a query.

The error_logging_clause is meant to declare the behavior of the task in case any errors occur.

ON COMMIT is the default schedule, therefore it can be omitted.

The CHANGES operator part of a continuous DML as above is implicitly assumed to specify the start with timestamp argument as LAST_SCHEDULE_TIME, so the above is equivalent to

```
CREATE TASK q [ON COMMIT] AS
INSERT INTO LogT
SELECT CONTINUOUS * FROM
CHANGES(T, LAST_SCHEDULE_TIME, 'DELTA');
```

LAST_SCHEDULE_TIME is a new expression that makes sense only in a continuous DML statement and that evaluates to the time of the last schedule for the task, if any exists, or to the time the task has been created otherwise. To be more precise, the semantics of LAST_SCHEDULE_TIME depends on the error_logging_clause. If the error_logging_clause is of the form

```
LOG ERRORS [INTO <table>(<simple_expression>)] REJECT LIMIT <limit_value> RETRY LIMIT
<retry_limit_value>
```

and if an error occurs during a scheduled execution of the task, and if the *limit_value* is reached due to the error, then LAST_SCHEDULE_TIME column is not updated.

We start with an empty table:

| 12:00> SELECT * FROM T; | 12:00> SELECT * FROM CHANGES(T, LAST_SCHEDULE_TIME, 'DELTA'); | 12:00> SELECT * FROM LogT; |
|---|---|---|

We insert 3 rows in T and then run a few statements in the same transaction:

| 12:01> INSERT INTO T VALUES ('A', 1), ('B', 2), ('C', 3); |
|---|

| 12:01> SELECT * FROM T; | 12:01> SELECT * FROM CHANGES(T, 12:00, 'DELTA'); | 12:01> SELECT * FROM LogT; |
|---|---|---|
| Key \| Val<br>A    1<br>B    2<br>C    3 | | |

No rows are returned from LogT because there are no committed changes on T since the creation at 12:00 of the task.

We now commit:

```
12:02> COMMIT;
```

| 12:02> SELECT * FROM T; | 12:02> SELECT * FROM CHANGES(T, 12:00, 'DELTA'); | 12:02> SELECT * LogT |
|---|---|---|
| Key \| Val<br>A    1<br>B    2<br>C    3 | Key \| Val \| Action \| RowID \| Time<br>A   1   INSERT   00000001   12:01<br>B   2   INSERT   00000002   12:01<br>C   3   INSERT   00000003   12:01 | Key \| Val \| Action \| RowID \| Time<br>A   1   INSERT   00000001   12:01<br>B   2   INSERT   00000002   12:01<br>C   3   INSERT   00000003   12:01 |

LAST_SCHEDULE_TIME in the context of the execution of the task becomes 12:02.

Note that we are assuming that Time is monotonically increasing with every COMMIT. If this is not a practical assumption, we can also define a COMMIT sequence column, say SYS$CommitNo, and rewrite our examples to include it.

Next, we update a row and delete another row, and finally commit:

```
12:03> UPDATE T SET Val = 20 WHERE Key='B';
12:04> DELETE FROM T WHERE Key = 'C';
12:05> COMMIT;
```

| 12:05> SELECT * FROM T; | 12:05> SELECT * FROM CHANGES(T, 12:02, 'DELTA'); | 12:05> SELECT * LogT |
|---|---|---|
| Key \| Val<br>A   1<br>B   20 | Key \| Val \| Action \| RowID \| Time<br>B   2    DELETE   00000002   12:03<br>B   20   UPDATE   00000002   12:03<br>C   3    DELETE   00000003   12:04 | Key \| Val \| Action \| RowID \| Time<br>A   1    INSERT   00000001   12:01<br>B   2    INSERT   00000002   12:01<br>C   3    INSERT   00000003   12:01<br>B   2    DELETE   00000002   12:03<br>B   20   UPDATE   00000002   12:03<br>C   3    DELETE   00000003   12:04 |

This example is akin to SQL Stream PUMP (CREATE PUMP :: SQLstream Documentation, see [27]).

### 5.2.1.2 Table Replication

We will use the same table T in the previous example. A table TR is created with Key, Val columns, together with RowID$ column that points to the row id in T and delta$ column that indicates the nature of the last operation in COMMIT order on the row. The goal is to keep TR in sync with T: TR.Key, TR.Val should contain the same committed data as T. For now, let's assume that no delay in this synchronization is allowed, to make the examples easier to read.

We can create a continuous task as shown in Figure 1:

Figure 1: Continuous replication of T into TR

```
12:00 > CREATE TASK q [ON COMMIT] AS

MERGE INTO TR USING (SELECT CONTINUOUS Key, Val, RowID RowId$, Action delta$

                     FROM CHANGES(T, [LAST_SCHEDULE_TIME,] 'DELTA')) S

ON (TR.RowID$=S.RowID$ AND (S.delta$ !=  'INSERT'))

WHEN MATCHED THEN UPDATE SET TR.Key = S.Key,

                      TR.Val = S.Val,

                      TR.RowID$ = S.RowID$,

                      TR.delta$ = S.delta$

              DELETE WHERE TR.delta$ = 'DELETE'

WHEN NOT MATCHED THEN INSERT (TR.Key, TR.Val) VALUES (S.Key, S.Val);
```

This is a special case of

```
CREATE TASK q [<schedule_clause>] AS
MERGE INTO  <table|insertable view>
```

```
USING <continuous_relation> ON <condition>
[<merge_update_clause>] [<merge_insert_clause>] [<error_logging_clause>]
```

where

```
<continuous_relation>  ::=
SELECT CONTINUOUS <select_list> FROM CHANGES(<relation>, <format>)
```

ON COMMIT is the default schedule, therefore it can be omitted.

Note that all tasks are actually continuous, see DBMS_SCEHDULER in Oracle ([20]) as well as [25], [27]. It may make more sense to use the CONTINUOUS token explicitly in the statement clause rather as a qualifier of the task. It is the CONTINUOUS token, wherever we agree to place it, that semantically differentiates these statements from simply creating a scheduled task that executes, at each schedule, a given statement. It controls how start time is computed in the evaluation of the CHANGES relational operator.

Each row in the target table TR for which the ON condition is true is updated, and, if the DELETE condition holds for the post update row image, the row is deleted. If the ON condition is not true for any rows, then the database inserts into the target table based on the corresponding source table row.

Note that it is essential for correctness that DELTA change format compacts the changes for the same row id. This is discussed in more detail in the next example.

We start with an empty table:

| 12:00> SELECT * FROM T; | 12:00> SELECT * FROM CHANGES(T, LAST_SCHEDULE_TIME, 'DELTA'); | 12:00> SELECT Key, Val FROM TR; |
|---|---|---|
| | | |

We insert 3 rows in T and run some statements in the same transaction:

| 12:01> INSERT INTO T VALUES ('A', 1), ('B', 2), ('C', 3); |
|---|

| 12:01> SELECT * FROM T;<br><br>Key \| Val<br>A    1<br>B    2<br>C    3 | 12:01> SELECT * FROM CHANGES(T, 12:00, 'DELTA'); | 12:01> SELECT Key, Val FROM TR; |
|---|---|---|

No rows are returned from TR because there are no committed changes on T since the creation at 12:00 of the task.

We now commit:

```
12:02> COMMIT;
```

| 12:02> SELECT * FROM T; | 12:02> SELECT * FROM CHANGES(T, 12:00, 'DELTA'); | 12:02> SELECT Key, Val FROM TR; |
|---|---|---|
| Key \| Val<br>A    1<br>B    2<br>C    3 | Key \| Val \| Action \| RowID \| Time<br>A    1    INSERT    00000001    12:01<br>B    2    INSERT    00000002    12:01<br>C    3    INSERT    00000003    12:01 | Key \| Val<br>A    1<br>B    2<br>C    3 |

LAST_SCHEDULE_TIME in the context of the execution of the task becomes 12:02.

Next, we update a row and delete another row, and finally commit:

```
12:03> UPDATE T SET Val = 20 WHERE Key='B';
12:04> DELETE FROM T WHERE Key = 'C';
12:05> COMMIT;
```

| 12:05> SELECT * FROM T; | 12:05> SELECT * FROM CHANGES(T, 12:02, 'DELTA'); | 12:05> SELECT Key, Val FROM TR; |
|---|---|---|
| Key \| Val<br>A    1<br>B    20 | Key \| Val \| Action \| RowID \| Time<br>B    2    DELETE    00000002    12:03<br>B    20    UPDATE    00000002    12:03<br>C    3    DELETE    00000003    12:04 | Key \| Val<br>A    1<br>B    20 |

(B, 2) in TR matches DELETE change record (B, 2, DELETE, 00000002, 12:03), therefore (B, 2) is deleted.
(C, 3) in TR matches DELETE change record (C, 3, DELETE, 00000003, 12:04), therefore (C, 3) is deleted.
(B, 20, UPDATE, 00000002, 12:03) does not match any of the rows in TR, therefore (B, 20) is inserted.

**Insert-Only Table Replication**

If relation T is known to be insert-only, then a simplified statement can be used:

```
12:00 > CREATE TASK q [ON COMMIT] AS
INSERT INTO TR
SELECT CONTINUOUS S.Key, S.Val
```

```
FROM CHANGES(T, [LAST_SCHEDULE_TIME,] 'DELTA')) S
```

### 5.2.1.3 Slow Changing Data

Consider now the SCD example. We can attempt to formalize it as in Figure 2.

---

Figure 2: Slow changing data

```
CREATE TASK q [ON COMMIT] AS
MERGE INTO T
USING
(SELECT CONTINUOUS <F join D columns>, RowID RowId$, Action delta$
FROM
CHANGES(SELECT * FROM F, D
        WHERE <join_condition>, 'DELTA')) S
ON (T.RowID$=S.RowID$ AND (S.delta$ != 'INSERT'))
WHEN MATCHED THEN UPDATE SET <T.key = S.key …>,
                            T.RowID$ = S.RowID$,
                            T.delta$ = S.delta$
                  DELETE WHERE T.delta$ = 'DELETE'
WHEN NOT MATCHED THEN INSERT (T.key…)  VALUES (S.key…);
```

---

The type of statement we write heavily depends on the semantics of the CHANGES operator, and it gets quite complex once we consider the possibility of LOG format as well.

If the change log format encodes UPDATE as DELETE followed by INSERT, then the correctness of the above MERGE statement depends on further assumptions, such as:

- Every row in the change log contains a RowID row globally unique identifier column
- Time timestamp of 2 rows committed separately cannot be the same, a.k.a timestamp increases with the COMMIT sequence

- UPDATE modeled as DELETE followed by INSERT maintains the RowID column value

- DELETE followed by INSERT for one logical UPDATE may have the same Time value

- Row ids are not recycled - every INSERT gets a new row id.

Under these simplifying assumptions, we still rely on further assumptions about the way MERGE statement is implemented:

- We need to order the source on RowID and Time, with ties ordered by 1st DELETE, then INSERT.
- We expect match and not matched actions to be taken in the ORDER in which rows are returned from source - the SQL execution will need to be serialized after the source computation.
- Matched DELETE rows are deleted, not matched INSERT rows are inserted.

A few optimizations are possible:
- GROUP source on RowID, order by Time, with ties ordered by 1st DELETE, then INSERT

- For each group, for pattern [INSERT] (DELETE INSERT)+, return the last INSERT in group order, for pattern * DELETE return the last DELETE in group order, for pattern * INSERT return the last INSERT in group order.

If the change log format represents UPDATE by a single record, then

- We need to order the source on RowID and Time.
- Matched DELETE rows are deleted, matched UPDATE rows are updated, not matched rows are INSERTED.

Similar optimizations are also possible:
GROUP source on RowID, order by Time
- For each group, for pattern INSERT (UPDATE)+, return INSERT with the row image as the post image of the last UPDATE in group order,
  - for pattern (UPDATE)+ return UPDATE with before image of first UPDATE and after image of last UPDATE rows in group order,
  - for pattern * DELETE return the last DELETE in group order, for pattern * INSERT returns the last INSERT in group order.

The basic questions that come up are:

- To express the requirement for a table T to be kept in sync with a statement/relation, should special short syntax be introduced, perhaps as a new continuous DML statement? In other words, should the SQL standard only express that T is kept in sync with a statement, in the worst case via full re-computation, and leave it to vendors to improve this computation, in the implementation? Or should users be expected to understand how to incrementally INSERT or MERGE change log records, under formats they control, for the same purpose?

- DELTA format CHANGES output is the result of a MERGE-like statement on the LOG format output. If we create special syntax for DELTA, should we not create special syntax for applying changes in general?

## 5.2.2 Apply Changes DML

These are simple examples and yet the MERGE statement still seems cumbersome. Imagine users having to type MERGE multiple times, for different tables/relations. This can become error prone and tedious.

The expert group discussed introducing a special APPLY CHANGES continuous DML, as

```
CREATE TASK q [<schedule_clause>] AS
APPLY CHANGES USING <continuous_relation>
TO <table|insertable view>
[<error_logging_clause>] [<initial_snapshot_clause>]
```

Using the above syntax, this is how EG would write the above examples:

- Change log:

```
CREATE TASK q [ON COMMIT] AS
APPLY CHANGES USING
SELECT CONTINUOUS *
FROM CHANGES(SELECT *
```

```
            FROM CHANGES(T, 'DELTA'),
            'DELTA')
TO LogT;
```

Note that CHANGES is a relational operator and can be nested, as in the example above.

- Table replication:

```
CREATE TASK q [ON COMMIT] AS
APPLY CHANGES USING
SELECT CONTINUOUS *
FROM CHANGES(T, 'DELTA')
TO TR;
```

- SDC:

```
CREATE TASK q [ON COMMIT] AS
APPLY CHANGES USING
SELECT CONTINUOUS *
FROM
CHANGES(SELECT * FROM F, D
      WHERE <join_condition>,
      'DELTA')
TO T;
```

- Top 10 stocks

```
CREATE TASK q [ON COMMIT] AS
APPLY CHANGES USING
SELECT CONTINUOUS *
FROM
CHANGES(SELECT *
      FROM Stocks
      ORDER BY <order_keys>
      DESC LIMIT 10,
      'DELTA')
TO T;
```

- Aggregation

```
CREATE TASK q [ON COMMIT] AS
APPLY CHANGES USING
SELECT CONTINUOUS *
FROM CHANGES(SELECT <grouping keys>,
                  COUNT(*),
                  MAX(S.val)
            FROM S
            GROUP BY <grouping keys>,
            'DELTA')
TO T;
```

### 5.2.3 Schedule Clause

Many databases have detailed and precise syntax for the schedule clause ([19], [24] etc.). EG believes that it is important to also include

- Periodic - lots of small details of the grammar
- `COMMIT`
- `COMMIT ASYNCHRONOUSLY`.

Optionally EG can consider

- `COMMIT ON <table|insertble_view>+ [ASYNCHRONOUS]`
- `DEMAND`
- end count or end time.

Combinations should also be allowed - for instance, PERIODIC EVERY 10 SECONDS ON COMMIT is a valid syntax for a schedule that should happen on every COMMIT of a transaction relevant to the statement, as well as every 10 seconds.

This clause can also control task overlap behavior - we can assume for now that all task executions are serialized, so essentially no overlap is possible.

### 5.2.4 Initial Snapshot

`WITH INITIAL SNAPSHOT`

If specified, all existing rows in the relation part of the continuous relation clause, as of the time the task is created, appear as INSERT change log records.

### 5.2.5 On Error Clause

Our suggestion is to use an error handling and logging clause and add to it specific syntax that makes sense for continuous DML, such as maximum retry count in case of error during the same task execution. For instance, we can start with

`LOG ERRORS [INTO <table>(<simple_expression>)] REJECT LIMIT <limit_value> RETRY LIMIT <retry_limit_value>`

### 5.2.6 Task State Machine

As expected, tasks being new SQL objects, DDLs will be introduced to manage them. Below is just an initial summary of possible DDLs and their effect of where changes are recomputed from as the task state changes. One key observation is that task state is persistent (checkpointed), and therefore maintained upon database restart.

|  | **To state** | **Last execution time** | **Notes** |
|---|---|---|---|
| **From state** |  |  |  |

|         | ACTIVE  | Time of COMMIT of the CREATE TASK DDL | Upon successful CREATE TASK DDL |
|---------|---------|---------------------------------------|----------------------------------|
| ACTIVE  | PAUSED  | unchanged                             | ALTER TASK q PAUSE;              |
| ACTIVE  | STOPPED | unchanged                             | ALTER TASK q STOP;               |
| PAUSED  | ACTIVE  | unchanged                             | ALTER TASK q RESUME;             |
| PAUSED  | STOPPED | unchanged                             | ALTER TASK q STOP;               |
| STOPPED | ACTIVE  | Time of COMMIT of ALTER TASK RESUME DDL | ALTER TASK q RESUME;           |
| STOPPED | PAUSED  | Not possible                          |                                  |
| Any     |         |                                       | DROP TASK q;                     |

# 6. Increasing Constraint – Definition, Usage and Generalization

Streaming solutions often imply that streaming sources have an obligatory timestamp column with monotonically increasing timestamp value upon each insert. Often, the timestamp represents system time, and therefore basic assumptions on clock synchronization in distributed settings are sufficient to guarantee monotonicity. It is common to also reason about application timestamps – these are defined outside of the ingestion system and therefore cannot be guaranteed or trusted as monotonically increasing. Most often, rows/events in streams may arrive out of order with respect to application time, though it is practical to expect some bounds on the out of order degree, or on how late events can either be accepted in the stream or, alternatively, considered for the purpose of processing.

Therefore EG starts from the basic observation that, to properly represent streaming processing in SQL, we must be able to reason about monotonicity of a column as well as about "almost" and bounded monotonicity, as we next detail.

## 6.1 Increasing (With Grace Period) Constraint

The expert group discussed extending the type of SQL constraints to include a new *monotonically increasing* constraint.

A monotonically increasing constraint guarantees that column values in a table can only increase. The constraint has two modes: STRICTLY INCREASING and INCREASING with corresponding semantics. It is applicable to columns with data types which accept the ">, <, <=, >=" operators. The (strictly) increasing constraint is equivalent to maintaining a max value for a column over the lifespan of a table, and verifying that new rows, or updated columns will always have value not less than (or equal to) this maximum.

The semantics of increasing constraint on column C with respect to DDL is as follows. Assume that the current maximum value of column C is C_MAX and constraint is STRICTLY INCREASING.

- Delete- Deleted rows must have C > C_MAX. The removal of data can be achieved using Expiration, which is elaborated in Section 8.

- Insert - The C_MAX will be updated to maximum(C_MAX, max(inserted_rows_C)). If multiple rows are inserted or loaded in a single statement, they do not have to be inserted in order, but all must have a value of C > C_MAX.

- Update - The value of C must be > C_MAX for all updated rows.

- Partition management operations - Dropping partition is analogous to Delete. Exchange or adding a partition is treated like insert, i.e., all new rows must have C value greater than C_MAX and after operation the C_MAX value is updated.

The syntax for the increasing constraint can be:

CONSTRAINT [STRICTLY ] INCREASING column-name <constraint state>

For example,

ALTER TABLE events CONSTRAINT INCREASING time;

[5] introduced a very useful notion of an INCREASING constraint with a grace period. The *grace period* defines an interval where we allow recent rows to arrive out of increasing order. Consider an application where multiple brokers report a bid for a house. The bids are recorded in the following table:

- CREATE TABLE bids (broker varchar(10), house varchar(30), tstamp date, bid_value number);

The following were records inserted into the table by different brokers:

('John Broker', 'SF Main St #1', '15-DEC-2019 14:55:00', 450,000)  - inserted at 14:55:00
 ('Andy Broker', 'SF Main St #1', '15-DEC-2019 14:55:10', 451,000)  - inserted at 14:55:10
('Bill Broker', 'SF Main St #1', '15-DEC-2019 14:55:20', 452,000)  - inserted at 14:55:20
 ('Mark Broker', 'SF Main St #1', '15-DEC-2019 14:55:30', 440,000)  - inserted at 14:55:30

Broker Mike received a bid for the house at '15-DEC-2019 14:55:05' but was not ready to insert till 14:55:50. Should his bid be accepted?

The grace period G allows us to accept inserts into a table with increasing constraint with pre-defined delay G. The idea is to accept values in the increasing column C between max( C) – G. If G = 30 seconds, Mike will be able to insert ('Mark Broker', 'SF Main St #1', '15-DEC-2019 14:55:05', 455,000) - inserted at 14:55:50. However if G = 10 sec, then Mike would not be able to insert his bid since we would insert up to 15-DEC-2019 14:55:20'. He is too late.

Grace periods are a special case of a concept known as a *watermark*. [5] defines watermark as follows: "if a watermark observed at processing time y has value of event time x, it is an assertion that as of processing time y, all future records will have event timestamps greater than x." This definition seems to imply that the event time has DATE data type. EG argues that increasing constraints should be applicable to any data type that can be ordered, as well as to DATE columns that represent application time rather than system time.

An increasing constraint with grace period can be defined as

CONSTRAINT [STRICTLY] INCREASING column-name GRACE <value> <constraint state>

For example,

ALTER TABLE events CONSTRAINT INCREASING time GRACE INTERVAL '20' second;

The constraint state allows us to specify if the constraint is DEFERRED or IMMEDIATE, or if it is ENABLED or DISABLED and if it is validated or not. Recall that ENABLE|DISABLE governs if the constraint is enabled or disabled on the new data. The DEFERRED | IMMEDIATE governs when an enabled constraint is checked. The former verifies it on commit, the latter at the end of a DML statement. The RELY records constraint in the metadata. Increasing constraints should support DEFERRED, IMMEDIATE, RELY, NONRELY, ENABLED and DISABLED options.

The constraint may not allow for VALIDATE | NONVALIDATE options (as we cannot check it in the past). When the constraint is turned to the ENABLED state, we get the maximum value of the column, and verify that all subsequent values will be greater than it.

We note that a long running transaction may have INSERTS, UPDATES and DELETES. If increasing constraint is DEFERRED, we will guarantee that at the end of the transaction all newly inserted or updated rows have C value greater than C_MAX. C_MAX will be updated at the end of the transaction.

## 6.2 Increasing Constraint-Aided Time-Windows

We illustrate the connection between increasing constraint and time-windows by means of a simple example. Consider an append-only relation *events* with the following rows, shown in COMMIT order:

| Event | Location | Time | measure |
|-------|----------|-----------|---------|
| A | SF | 15-NOV-19 | 30.00 |
| B | SF | 16-NOV-19 | 32.00 |
| A | SF | 17-NOV-19 | 30.00 |
| A | SF | 18-NOV-19 | 31.00 |
| B | SF | 19-NOV-19 | 34.00 |
| A | SF | 20-NOV-19 | 30.00 |
| B | SF | 21-NOV-19 | 33.00 |
| A | SF | 22-NOV-19 | 35.00 |
| A | SF | 26-NOV-19 | 35.00 |
| B | SF | 27-NOV-19 | 34.00 |
| A | SF | 28-NOV-19 | 33.00 |
| B | SF | 29-NOV-19 | 34.00 |
| B | SF | 30-NOV-19 | 34.00 |

*Events* has an *event* column, a *location* column, a *measure* numeric column, and a *time* column as DATE. In this example *time* values are increasing. We want to compute every second calendar day the sum of *measure* for the last 5 days, per *location*.

Following the type of syntax of [5], we would formulate this as

```
SELECT location, win_start, win_end, sum(measure) sum_m
FROM
events WINDOW(
	time
	START_WITH '15-NOV-19'
	RANGE INTERVAL '5' DAY
	ADVANCE INTERVAL '2' DAY
	BOUNDS (win_start, win_end))
GROUP BY location, win_start, win_end;
```

The WINDOW clause over the events table:

```
WINDOW(
        time
        START_WITH '15-NOV-19'
        RANGE INTERVAL '5' DAY
        ADVANCE INTERVAL '2' DAY
        BOUNDS (win_start, win_end))
```

specifies

- the timestamp column (*time*)

- the starting time of the following windows: 15-NOV-19 00:00:00

- the length of the window (RANGE INTERVAL '5' DAY)- this is specified using INTERVAL type

- the advancement of the window (ADVANCE INTERVAL '2' DAY) - specified using INTERVAL type

- new columns that define window start and window end (BOUNDS (win_start, win_end)) clause.

A table with a WINDOW clause will be called a *windowing table*.

The above would be formulated in KSQL as

SELECT location, sum(measure) sum_m

FROM

events WINDOW HOPPING(SIZE '5' DAYS,

          ADVANCE BY '2' DAYS)

GROUP BY location

EMIT CHANGES;

Given the data in *events*, we have the following time windows produced by the query:

- window 1 starting 15-NOV-19 and ending 19-NOV-19. Total of 5 rows. Window size 5 days.

- window 2 starting 17-NOV-19 and ending 21-NOV-19. Total of 5 rows. Window size 5 days.

- window 3 starting 19-NOV-19 and ending 23-NOV-19. Total of 4 rows. Window size 5 days.

- window 4 starting 21 -NOV-19 and ending 25-NOV-19. Total of 2 rows. Window size 5 days.

- window 5 starting 23-NOV-19 and ending 27-NOV-19. Total of 2 rows. Window size 5 days.

- window 6 starting 25-NOV-19 and ending 29-NOV-19. Total of 4 rows. Window size 5 days.

- window 7 starting 27-NOV-19 and ending 01-DEC-19. Total of 4 rows.

- window 8 starting 29-NOV-19 and ending 03-DEC-19. Total of 2 rows.

Since there are multiple rows in each window, and we assume that each row inserted and committed, on COMMIT query execution would generate corrections to previously reported windows. Therefore EG proposes to make this semantics clear and express the statement as a continuous cursor:

```
CREATE CONTINUOUS CURSOR c as

SELECT Action, location, win_start, win_end,  sum(measure) sum_m

FROM

CHANGES(
SELECT location, win_start, win_end, sum(measure) sum_m
FROM
events WINDOW(
            time
            START_WITH '15-NOV-19'
            RANGE INTERVAL '5' DAY
            ADVANCE INTERVAL '2' DAY
            BOUNDS (win_start, win_end))

GROUP BY location, win_start, win_end);
```

Note that we purposefully specify the window bounds as grouping keys, since there should not be a need for implicit grouping.

The following rows will be fetched from the above continuous cursor, in order:

- (+, SF, 15-NOV-19, 19-NOV-19, 30)

- (-, SF, 15-NOV-19, 19-NOV-19, 30), (+, SF, 15-NOV-19, 19-NOV-19, 62)

- (-, SF, 15-NOV-19, 19-NOV-19, 62), (+, SF, 15-NOV-19, 19-NOV-19, 92), (+, SF, 17-NOV-19, 21-NOV-19, 30)

- (-, SF, 15-NOV-19, 19-NOV-19, 92), (+, SF, 15-NOV-19, 19-NOV-19, 121), (-, SF, 17-NOV-19, 21-NOV-19, 30), (+, SF, 17-NOV-19, 21-NOV-19, 61)

- (-, SF, 15-NOV-19, 19-NOV-19, 121), (+, SF, 15-NOV-19, 19-NOV-19, 154), (-, SF, 17-NOV-19, 21-NOV-19, 61), (+, SF, 17-NOV-19, 21-NOV-19, 95), (+, SF, 19-NOV-19, 23-NOV-19, 34)

and so on. The final 2 rows fetched will be (-, SF, 29-NOV-19, 03-DEC-19, 34), (+, SF, 29-NOV-19, 03-DEC-19, 68). Notice that, for instance, no rows for the 1st window will be reported after the rows shown above.

This is very verbose. The user might instead be interested in learning just once about each window. If

```
ALTER TABLE events CONSTRAINT INCREASING time;
```

was issued before the first fetch from the cursor, then reporting each window once no further rows are possible to be added to the window is possible – as soon as a row is inserted into *events* with time above the window's upper bound, the window can be reported. To indicate this behavior, new syntax should be present, and the EG considered using a FINAL keyword, as:

```
CREATE CONTINUOUS CURSOR c as

FINAL(
      SELECT Action, location, win_start, win_end,  sum(measure) sum_m
      FROM
      CHANGES(
            SELECT location, win_start, win_end, sum(measure) sum_m
            FROM
            events WINDOW(
                  time
                  START_WITH '15-NOV-19'
            RANGE INTERVAL '5' DAY
                  ADVANCE INTERVAL '2' DAY
                  BOUNDS (win_start, win_end))
      GROUP BY location, win_start, win_end));
```

The following rows will be fetched from the above continuous cursor, in order:

- (+, SF, 15-NOV-19, 19-NOV-19, 157)

- (+, SF, 17-NOV-19, 21-NOV-19, 158)

- (+, SF, 19-NOV-19, 23-NOV-19, 132)

- (+, SF, 21-NOV-19, 25-NOV-19, 68)

- (+, SF, 23-NOV-19, 27-NOV-19, 69)

- (+, SF, 25-NOV-19, 29-NOV-19, 136)

No rows for the last 2 windows are reported, under the proposed semantics for FINAL, due to the presence of the INCREASING constraint. In this case, it is valid to remove CHANGES operator and simply define

```
CREATE CONTINUOUS CURSOR c as

FINAL(
SELECT location, win_start, win_end,  sum(measure) sum_m
FROM
events WINDOW(
            time
            START_WITH '15-NOV-19'
```

```
                RANGE INTERVAL '5' DAY
                ADVANCE INTERVAL '2' DAY
                BOUNDS (win_start, win_end))
```

GROUP BY location, win_start, win_end);

The above syntax is akin to KSQL (see [14]) syntax

SELECT location, sum(measure) sum_m

FROM

events WINDOW HOPPING(SIZE '5' DAYS,

        ADVANCE BY '2' DAYS)

GROUP BY location

EMIT FINAL;

Alternatively, assume that an increasing with grace period constraint is enabled on the relation:

`ALTER TABLE events CONSTRAINT INCREASING time GRACE INTERVAL '1' DAY TO SECOND;`

Then the following rows are fetched from the last cursor:

- (+, SF, 15-NOV-19, 19-NOV-19, 157)

- (+, SF, 17-NOV-19, 21-NOV-19, 158)

- (+, SF, 19-NOV-19, 23-NOV-19, 132)

- (+, SF, 21-NOV-19, 25-NOV-19, 68)

- (+, SF, 23-NOV-19, 27-NOV-19, 69)

No rows are returned for a window with bounds 25-NOV-19, 29-NOV-19 as no date above 30-NOV-19 exists in *relation* yet.

Increasing with grace period constraint formalizes a particular type of constraint on write – for append-only tables, it allows insertion of events that are either monotonically increasing in the time dimension or at most lagging behind by a fixed period.

The grace period is a constant. If the grace period needs to change – say to accommodate a larger delay in receiving events -, then the user needs to ALTER the relation and specify the new grace period. If the grace period needs to be reduced however, then the cursors also need to be invalidated and computation reinstated.

It is not always possible to prevent the insertion of rows. For instance, if we replay a stream of events and would like to compute an aggregate over a sliding time window for an application time dimension, there is no opportunity to discard events. In this case, we would like to still offer the user the option to select the final row per window, rather than subscribe to the verbose delta-semantics cursor.

## 6.3 Window Subquery

The EG discussed standardizing *window subquery*

```
relation WINDOW (windowing_column

    [START WITH start_time]

    RANGE rane_interval

    [GRACE grace_period]

    [ADVANCE advance_interval]

    [BOUNDS (start_bound, end_bound)])
```

- *windowing column*: column in *relation*, either of DATE/TIMESTAMP or numeric data type

- *start_time*, *range_interval*, *advance_interval* and *grace_period* of data types compatible with the *windowing_column* and must evaluate to constant expressions at query compile-time

- when *start_time* is not specified, use 0 for numeric and a system-level time constant otherwise

- when not specified, *advance_interval* is *range_interval*

- *start_bound* and *end_bound* must be column names (identifiers), for defining the names of the window start and end columns; when not specified, WIN_START and WIN_END are used

- Note: EG proposes to keep "window" in the name of the clause, as users are accustomed to. To differentiate it from existing SQL windows, we can refer to the above as a window *subquery*, as it can appear in a FROM clause.

Window subquery in a cursor defined as FINAL has the following semantics:

- Applies to a relation and returns another relation.

- It logically partitions the input relation into ranges.

- WINDOW (C START WITH S RANGE R ADVANCE A …) ranges are [S, S+R),…, [S+A*I, S+A*I + R), … .

- Ranges are non overlapping (*advance_interval* >= *range_range_interval*) or overlapping otherwise.

- The clause returns all columns of relation, plus *start_bound* and *end_bound*, for range bounds.

- An input row can be mapped to none, one or multiple (finitely many) ranges defined by the clause.

- All ranges starting before *start_time* are assumed "closed".

- When part of a continuous cursor, a range becomes "closed" at the end of each cursor fetch. Once a range is "closed", it remains "closed".

- Therefore, rows mapped to a "closed" range present in *relation* at a subsequent fetch are not returned by the clause.

- If *grace_period* is specified, then a range [ws, we) becomes "closed" if there is a row r in r*elation* above the grace period – with r.windowing_column > we + grace_period.

- If no *grace_period* is specified and a constraint INCREASING WITH GRACE *column_grace_period* is enabled on *relation.windowing_column*, then *column_grace_period* will be used in the same way.

- It should be feasible to answer "Is range [ws, we) "closed" ?" without requiring unbounded storage.

- One reasonable condition to guarantee bounded storage is that *relation* must be a table or view with enabled constraint INCREASING [WITH GRACE *column_grace_period*] on *relation.windowing_column*. Clauses can still specify their own *grace_period*, to take precedence.

- Invariant: No value < (SELECT MAX(*windowing_column*) - *column_grace_period* FROM *relation*) can exist in *relation.windowing_column*.

• Invariant: All ranges [ws, we) with we < (SELECT MAX(windowing_column) FROM relation) are "closed".

• We keep state only for the remaining ranges.

## 6.4 Note on Hop and Tumble Windows

As an exercise, we show by example that a combination of continuous DML and continuous cursor can have the same semantics as hop and tumble window subqueries in FINAL statements. However, the rewrite is sufficiently complex to demonstrate the value of special syntax. We include the example only as a justification for maintaining special syntax for hop and tumble window subqueries.

Consider for instance

```
CREATE CONTINUOUS CURSOR c AS

SELECT FINAL eventHour, productId,

           COUNT(*) as orderCount

FROM

Orders WINDOW (

   eventTime

   RANGE INTERVAL '1' HOUR
```

```
    [ADVANCE INTERVAL '1' HOUR ]
    GRACE INTERVAL '10' SECONDS
    BOUNDS (eventHour, win_end))
GROUP BY eventHour, productId;
```

We can alternatively define one continuous task and one continuous cursor, as shown in Figure 3.

Figure 3: Tumbling window rewrite

```
CREATE CONTINUOUS TASK myTask AS
INSERT INTO t VALUES(eventHour, productId, orderCount)
SELECT FLOOR(eventTime TO HOUR) as eventHour, productId, COUNT(*) as orderCount
FROM Orders
WHERE EXISTS (SELECT 1 FROM Orders o2
            WHERE o2.eventTime >= FLOOR(eventTime TO HOUR) + INTERVAL '10' SECONDS)
            -- grace period expires by virtue of a row inserted above the grace period
                AND NOT EXIST (SELECT 1 FROM t WHERE t.eventHour = eventHour)
            -- search for history of closed ranges
GROUP BY eventHour, productid;
```

and

```
CREATE CONTINUOUS CURSOR c AS
FINAL(SELECT eventHour, productId, orderCount
      FROM t);
```

Similarly, consider

```
CREATE CONTINUOUS CURSOR c AS
FINAL(
SELECT win_start, win_end,
           productId,
           COUNT(*) as orderCount
FROM
Orders WINDOW(
   eventTime
   RANGE INTERVAL '1' HOUR
   ADVANCE INTERVAL '10' MINUTES
   GRACE INTERVAL '10' SECONDS
   BOUNDS (eventHour, win_end))
GROUP BY win_start, win_end, eventHour, productId);
```

We can alternatively rewrite as in Figure 4.

Figure 4: Hopping window rewrite

```
CREATE CONTINUOUS TASK myTask AS
INSERT INTO t VALUES(gid, productId, orderCount)
FINAL(SELECT g.COLUMN_VALUE as gid, -- unique id of the range
           o.productId, COUNT(*) as orderCount
FROM (Orders o, TABLE(range_identifiers(o.eventTime,
                                        INTERVAL '1' HOUR ,
                                        INTERVAL '10' MINUTES )) g
          -- left correlation join to generate rows in Orders mapped to ranges of eventTime
```

```
WHERE EXISTS (SELECT 1 FROM Orders oi

              WHERE oi.eventTime >= FLOOR(o.eventTime TO HOUR) + INTERVAL '10' SECONDS)

              -- grace period expires by virtue of a row inserted above the grace period

              AND NOT EXIST (SELECT 1 FROM t WHERE t.gid = g.gid)

                -- search for history of closed ranges

GROUP BY gid, productid);
```

where range_identifiers is a function that returns the array of range identifiers the 1st argument maps to, given that the 2nd argument is the range of the window, and the 3rd argument is the advance / width range value.

```
CREATE CONTINUOUS CURSOR c AS

FINAL(SELECT start_time(gid, INTERVAL '10' MINUTES) as win_start,

                    -- scalar computation of range start

              end_time(gid, INTERVAL '10' MINUTES, INTERVAL '1' HOUR) as win_end,

                    -- range end

          productId, orderCount

FROM t);
```

In short

- EG argues in favor of a simple clause for expressing both hop and tumbling final windows with a grace period.

- The clause is a table function and can be present anywhere where a table function is allowed.

- EG proposes standardizing INCREASING [GRACE] constraint on columns.

# 7. Finalization

One direction the group considered is to generalize insert-only tables and increasing constraints into a single concept: *finalization*. Existing systems such as Apache Flink and Apache Beam support programmable watermarks, which suggests that other kinds of constraint would be useful. Finalization allows the expression of insert-only, increasing, increasing-with-grace-period, and programmable watermarks in terms of boolean predicates. It also enables more exotic concepts like percentile watermarks and multi-dimensional watermarks.

The basic idea of finalization is that the programmer provides a boolean predicate on a table which determines which rows in the relation are final. Final rows in the relation cannot be updated or deleted, and new rows for which the finalization predicate holds cannot be inserted. Finalization predicates can vary over time, for example by making reference to `current_timestamp` or containing a subquery to another table. However, they must also be monotonic with respect to time. That is, once the predicate is `true` for a given row, it must be `true` for that row at all later times. Otherwise, the guarantee that a row is "final" would not hold.

Finalization predicates can take two forms:

1. A constraint on a persistent base table (AKA "finalization on write"), which prevents DML statements from modifying final rows in the table by aborting the transaction. For example, insert-only tables finalize rows immediately after they are inserted. An increasing constraint on a column `C` finalizes all rows where `C` is less than the current maximum in the table.
2. A clause in a query expression (AKA "finalization on read"), which applies the finalization predicate to the resulting table without affecting the sources of the data. For example, grace periods, which are usually applied ad hoc as part of a streaming query, can be viewed as an increasing constraint on a query expression, where they need not be part of a specialized `WINDOW` clause.

We present several examples to illustrate the use of finalization.

First, consider an accounting ledger. The accounting department will typically "close the books" once all entries up to a certain date have been recorded. Some consumers of the ledger wish to read entries while they're still subject to change, while others wish to only see entries after books are closed. This pattern can be implemented with finalization:

```
CREATE TABLE ledger(
     time timestamp,
     account integer,
     credit number);

-- Close the books for October:
ALTER TABLE ledger FINALIZE WHERE time < '2023-11-01';

SELECT * FROM ledger; -- Query all ledger entries.

FINAL(SELECT * FROM ledger); -- Query only "closed book" entries.
```

Next consider a stream processing use case, where we have a stream of events and wish to compute a windowed sum over some measure. The watermark for this stream of events is provided as another stream of timestamps—when a watermark is received, it indicates that all events with timestamp less than or equal to it have been received. These two streams can be combined using finalization to produce a windowed aggregate that only produces rows once they are complete.

```
CREATE TABLE events(
     metric text,
     time timestamp,
     measure number,
     INSERT ONLY);

CREATE TABLE watermarks(
     wm timestamp,
     INSERT ONLY);
```

```
WITH finalized_events AS (
    SELECT *
    FROM events
    FINALIZE WHERE time <= (SELECT max(wm) FROM watermarks))

FINAL(
    SELECT metric,
           date_trunc(minute, time) minute,
           sum(measure) sum_m
    FROM finalized_events
    GROUP BY metric, minute);
```

These examples were chosen because neither can be expressed as insert-only or increasing constraints. The main advantage of finalization is its generality, which makes it useful in a larger variety of situations without requiring the database implementation to be extended. In general, streaming often requires dealing with data which arrive out-of-order, and finalization seeks to address this problem holistically.

The main disadvantage of finalization, as acknowledged by the group, is its potential for implementation complexity and confusion among users. Finalization is a novel concept not yet available in existing products, so both its feasibility and practical utility are theoretical. It seems premature to standardize such a concept, so the group deemed it appropriate to leave such development to implementers of streaming systems.

# 8. Expiration

Time-to-live policies are common practice in industry (see [11] ) and of recurrent academic interest (see [21] for a recent example). Streaming SQL standardization effort has the timely opportunity and duty, as some see, to include declarative time-to-live language definition.

The group focused on formalizing a notion of *expiration*, as the ability to automatically remove rows from a table and mark them as *expired*. This can be thought of as "forgetting" a row without logically removing it from the table. The row can be physically removed from storage, or *purged*, asynchronously after being expired.

Expired rows are not accessible to new queries but are still logically part of the table. For instance, the CHANGES operator would ignore expired rows and not return them as deleted. Similarly, rows constrained to be final (and thus not able to be deleted) can be expired without affecting the constraint. Once a row is expired, it is eligible to be purged in the background at a future point in time. Therefore, there is a duration wherein a row can be expired but not physically removed from the table.

An alternative for supporting expiration is by making it explicit. Users choose when to expire the data in tables based on monotonic, deterministic at a point-in-time expressions that they define with respect to these tables. These expressions can include AND / OR clauses, reference multiple columns in the source, and sub-queries. Vendors are free to enforce which predicates are supported based on underlying implementation.

An expression can be specified on a persistent table at creation time. Rows matching that expression would expire when they match that expression. To facilitate ease of use we envision only allowing one expiration policy on a stream or table at a time. If the policy needs to be extended it can be modified via the ALTER DDL as shown below.

```
CREATE TABLE <name> EXPIRE WHERE <expression>
```

For example, a debug log may be expired daily:

```
CREATE TABLE LOGS (level varchar, message varchar, ts timestamp)
EXPIRE WHERE
      (level = 'DEBUG' and ts < dateadd(-1, 'day', now()));
```

An expression could be specified on a table after it has been created. Rows matching that expression would expire when they match that expression. EG envisions that the standard should not impose any restrictions on the predicates associated with the existing expiration clause during ALTER although EG believes that we should provide recommendations on what should be accepted during an ALTER statement, namely more strict predicates than what is already defined, as it can affect subsequent queries on the table.

```
ALTER TABLE <name> {ADD | MODIFY} EXPIRE WHERE <expression>
```

As an example, we modify our existing expiration policy to account for WARN level logs and expire them after one week:

```
ALTER TABLE LOGS
MODIFY EXPIRE WHERE
      (level = 'DEBUG' and ts < dateadd(-1, 'day', now())
      OR (level = 'WARN' and ts < dateadd(-7, 'day', now()));
```

If users wish to remove an expiration policy, they can simply unset the expiration expression:

```
ALTER TABLE <name> DROP EXPIRE;
```
such as
```
ALTER TABLE LOGS DROP EXPIRE;
```

With expiration defined on a table, regions of the table will have three states that the rows can be in:

- *not expired*: In this state the rows are active within a table

- *expired*: The rows are no longer relevant as determined by the expiration policy defined on a table. Rows can transition from the "Not Expired" State to this one although one can insert already-expired data into a table.

- *purged*: The rows are physically removed from the data system and are no longer accessible to new queries. Rows only transition to this state from the "Expired" State.

If a query targets expired data there are two primary ways this can be handled:

- *Silently Ignoring the Expired Rows*. The system could choose to allow queries whose results depend on expired data. This could be misleading to users as it gets hard to distinguish rows that were never present in the table based on the expression vs. rows that did match the predicate but were expired. For instance, consider the case of a query that counts records 15 days into the past, but expiration occurs at 14 days. If an event occurs between day 14 and 15 and gets automatically filtered out, the user will see a misleading count and may think that the event never occurred.

- *Throwing an Exception***:** If expired data is selected as part of the query the system could choose to throw an exception informing the user that expired data would be selected and that they should refine their queries which could include the relevant expiration expression based on the user's privileges. This allows the user to more safely reason about their query result sets. However, if the user doesn't care about a particular range of expired data it may inadvertently block the query.

The EG considered whether the user must opt-in if they want to factor in expired rows in their queries and throw an exception if expired rows could affect the result. Calculating whether a row was expired can increase the overhead of a query and impact users that have no idea that expiration is even defined on a table.

If a user has opted into expiration during queries via the use of select final then querying the table will result in the following:
- Only querying Not-Expired rows will result in all rows being returned
- Querying rows in the Not-Expired and Expired regions will result in an error
- Querying rows only in the Expired region will result in an error
- Querying rows in the Expired and Purged regions will result in an error
- Querying rows that span all regions will result in an error

The EG only briefly considered details on the interactions between expiration, transactions, referential integrity, and triggers. An in-depth investigation is left as future work.

# 9. Conclusions

Absent an official final report from the INCITS Data Management Expert Group (EG) on SQL Support for Streaming Data, this paper tries to summarize the most important topics considered by the EG. The EG encourages resumption of these standardization efforts upon further validation of the feasibility and utility of the above topics in the market.

# 10. Acknowledgments

EG would like to thank those who have participated in system demos and presentations, discussions, brainstorming, and various manuscripts drafting. Including: Tyler Akidau, Ken Chen, Florian Eiden, Amit Gurdasani, Richard Hillegas, Fabian Hueske, Julian Hyde, Tyler Jones, Krishna Kulkarni, Frank McSherry, Jim Melton, Daniel Millis, Shailendra Mishra, Sabina Petride, Jeff Pollock, Alex Reizman, Matthias Sax, Nigel Thomas, Prabhu Thukkaram, Alexey Leonov-Vendrovskiy, Timo Walther, Guozhang Wang, Calisto Zuzarte.

# 11. References

[1] A. Arasu, S. Babu, J. Widom: An Abstract Semantics and Concrete Language for Continuous Queries Over Data Streams. DBPL 2003

[2] Tyler Akidau, Paul Barbier, Istvan Cseri, Fabian Hueske, Tyler Jones, Sasha Lionheart, Daniel Mills, Dzmitry Pauliukevich, Lukas Probst, Niklas Semmler, Dan Sotolongo, and Boyuan Zhang, What's the Difference? Incremental Processing with Change Queries in Snowflake, Proc. ACM Manag. Data 1, 2, Article 196 (June 2023), 27 pages. https://doi.org/10.1145/3589776

[3] Azure Cosmos DB, Time-to-live in Azure Cosmos DB, Expire data in Azure Cosmos DB with Time to Live | Microsoft Learn

[4] S E. Begoli, T. Akidau . Suggestions for Extensions to SQL Standard to Support Uniform SQL Syntax for Streaming and Relational Structures. April 24, 2019

[5] Edmond Begoli, Tyler Akidau, Fabian Hueske, Julian Hyde, Kathryn Knight, and Keneth L. Knowles, One SQL to Rule Them All – an Efficient and Syntactically Idiomatic Approach to Management of Streams and Tables, SIGMOD 2019, 1757-1992, https://doi.org/10.1145/3299869.3314040

[6] D. Carney, U. Cetintemel, M. Cherniack, C. Convey, S. Lee, G. Seidman, M. Stonebraker, N. Tatbul, S. B. Zdonik: Monitoring Streams - A New Class of Data Management Applications. VLDB 2002

[7] S. Chandrasekaran, O. Cooper, A. Deshpande, M. J. Franklin, J. M. Hellerstein, W. Hong, S. Krishnamurthy, S. Madden, V. Raman, F. Reiss, M. A. Shah: TelegraphCQ: Continuous Dataflow Processing for an Uncertain World. CIDR 2003

[8] J. Chen, D. J. DeWitt, F. Tian, Y. Wang: NiagaraCQ: A Scalable Continuous Query System for Internet Databases. SIGMOD Conference 2000

[9] C. D. Cranor, T. Johnson, O. Spatscheck, V. Shkapenyuk: Gigascope: A Stream Database for Network Applications. SIGMOD Conference 2003: 647-651

[10] DynamoDB, Expiring items by using DynamoDB Time to Live, Expiring items by using DynamoDB Time to Live (TTL) - Amazon DynamoDB

[11] Google, About time to live, About time to live (TTL) | Cloud Spanner | Google Cloud

[12] INCITS – https://www.incits.org

[13] InfluxDB: MIT – open source stream database https://docs.influxdata.com/influxdb/v1.7/query_language/

[14] KSQL - https://docs.ksqldb.io

[15] Y. N. Law, H. Wang, C. Zaniolo: Query Languages and Data Models for Database Sequences and Data Streams. VLDB 2004

[16] Materialize, Temporal Filters: Enabling Windowed Queries in Materialize, Temporal Filters: Enabling Windowed Queries in Materialize

[17] Materialize, Temporal filters (time windows), Temporal filters (time windows) | Materialize Documentation

[18] J. Michels, K. Hare, K. Kulkarni, C. Zuzarte, Z. H. Lui, B. Hammerschmidt and F. Zemke, The new and improved SQL: 2016 Standard, ACM Sigmod Record 47(2), 2018, 51-61

[19] Oracle, Calendaring syntax for DBMS_SCHEDULER in Oracle

[20] Oracle, DBMS_SCHEDULER (oracle.com)

[21] S. Sankar and M. Athanassoulis, Query Language Support for Timely Data Deletion, in Proceedings of the 25[th] International Conference on Extending Database Technologies, Vol 2, pages 429-434, 2022

[22] Sankar Subramanian, Srikanth Bellamkoda, Huagang Li, Vince Liang, Lei Sheng, Wayne Smith, James Terry, Andrew Witkowski, Continuous Queries in Oracle, VLDB 2007, 1173-1184

[23] Snowflake, Continuous Data Pipeline Examples | Snowflake Documentation

[24] Snowflake, CRON-expression for Snowflake tasks

[25] Snowflake, Introduction to Tasks | Snowflake Documentation)

[26] ISO/IEC 9075-1:2023 - Information technology — Database languages — SQL — Part 1: Framework (SQL/Framework)

[27] SQLStreams, Pumps, Streams, and Processing Pipelines :: SQLstream Documentation